\documentclass{rspublic}
\usepackage{amsmath}
\usepackage{amsfonts}
\usepackage{amssymb}
\usepackage[Symbol]{upgreek}
\usepackage[nointegrals]{wasysym}
\usepackage{mathrsfs}
\usepackage{graphicx}
\newcommand{\bfu}{{\bf u}}
\newcommand{\bfx}{{\bf x}}
\newcommand{\Upo}{{\Sigma}}
\newcommand{\bdy}{{\partial \Sigma}}
\begin{document}
\title[Planar Navier-Stokes Equations in Bounded Domain]{\large Viscous flow regimes in a square. Part 2. Impact and rebound process of vortex dipole-wall interaction}
\author[F. Lam]{F. Lam}
%
%
\label{firstpage}
\maketitle
\begin{abstract}{Planar Navier-Stokes Equations; Vorticity; Stream Function; Non-linearity; Laminar Flow; Transition; Turbulence; Diffusion}

In this technical note, we demonstrate the robustness of our numerical scheme of vorticity iteration in dealing with the dipole-wall interaction at small viscosity, with emphasis on mesh convergence, boundary vorticity as well as wall viscous dissipation. In particular, it is found that, among the four different dipole configurations, the processes of vortex-wall collision at no-slip surfaces are exceedingly complex and are functions of the initial conditions. The critical issue direct numerical simulations is to establish mesh convergence which appears to be case-dependent. Roughly speaking, converged $2D$ meshes are found to be inversely proportional to viscosity at uniform spacings. Essentially, we have ruled out the possibility of anomalous energy dissipation in the limit of small viscosity. Our computational results show that the rate of the energy degradation follows the predictive trend of the well-known Prandtl scaling.
\end{abstract}
%
%
%
\section{Background}\label{intro}
The computation of viscous flows in the presence of solid surfaces is a delicate matter in numerical analysis. Consider the equations of motion of incompressible flow in $2$ space dimensions: 
\begin{equation} \label{ns}
	\nabla.\bfu = 0, \;\;\; \partial_t \bfu  + \nu \nabla{\times}\zeta = - (\bfu. \nabla ) \bfu  - \nabla p,
\end{equation}
where $\bfu{=}(u,v)$ denotes the velocity, and $p$ the pressure (unit density) which are treated as a continuum. All symbols have their usual meanings in fluid dynamics. In the vorticity-stream function formulation, the dynamics is described by
\begin{equation} \label{vort}
\Delta \psi = - \zeta,\;\;\; \partial_t \zeta  - \nu \Delta  \zeta = - {\partial_y \psi} \; {\partial_x \zeta} + {\partial_x \psi} \; {\partial_y \zeta}.
\end{equation}
The no-slip condition $\bfu_{\bdy} = 0$ applies for $t \geq 0$ on the four sides of the unit square, and this boundary condition implies 
\begin{equation*}
	\psi_{\bdy} = 0.
\end{equation*}
Once the stream function $\psi$ is calculated for known $\zeta$, the velocity is recovered as $(u, v) = ( \partial_y \psi, \; -\partial_x \psi )$. We are mainly interested in the transient Navier-Stokes dynamics from given initial solenoidal data $\bfu_0$ or in terms of vorticity $\zeta_0=\nabla{\times}\bfu_0$. At $t=0$, the solenoidal $\bfu_0$ may also be recovered from $\psi_0$ or $\zeta_0$.

The pressure Poisson equation is solved to obtain the pressure gradients
\begin{equation} \label{modp}
\Delta p = 2 \: \big( \partial_x u \: \partial_y v - \partial_y u\: \partial_x v \big), 
\end{equation}
subject to the Neumann boundary conditions, $\partial_x p$ and $\partial_y p$, which are obtained from (\ref{ns}) for known vorticity and velocity for $t>0$. 
In practice, the pressure at the start $t=0$ is somehow unspecified because $\partial_t \bfu_0$ is not available (unless assumed otherwise). Nevertheless, the initial data may be mathematically assigned as a step function $\bfu=0,\;t<0$, and $\bfu=\bfu_0, \; t \geq 0$. In this theoretical setting, the {\it initial pressure gradients} may be fixed in terms of generalised functions according to the momentum equations.

It is known that the vorticity evolves in a self-contained manner. Our numerical procedure is to determine the fixed-point solutions ($\zeta, \bfu$) at given time $t$. This can be done efficiently by an iteration procedure (Lam 2018). To solve the vorticity dynamics numerically, the unit square is subdivided into equally-spaced grids, denoted by $n$, and the grid points by ($i,j$). We use the implicit Euler scheme for time discretisation and a semi-implicit scheme for the non-linear term. Let $k$ denote the time step. The discretised vorticity matrix (size $n^2$) is iterated until the error difference satisfies a prescribed convergence criterion
\begin{equation} \label{cv}
	\delta \Pi=\sum_{ij} \Big| \;\Pi^{k+1}-\Pi^{k}\; \Big| < \epsilon.
\end{equation}
The difference $\delta \Pi$ may be scaled by the previous error size if $|\Pi^{k}| > 1$. Throughout the present calculations, we set the tolerance $\epsilon=10^{-8}$. As a general tool, no symmetry conditions have been imposed in our implementation. In what follows we will make an effort to examine the problem of mesh convergence with attention to the simulations of dipole-wall head-on impact. We will hence investigate how flow energy is redistributed and dissipated at different viscosities.
\section{Energy dissipation}
The energy and enstrophy are defined by
\begin{equation*} 
	E(t)=\frac{1}{2} \int_{\Upo} \big( u^2 + v^2 \big)(\bfx,t) \; \rd \bfx, \;\;\;\mbox{and}\;\;\; \Omega(t)=\int_{\Upo} \zeta^2 (\bfx,t) \; \rd \bfx,
\end{equation*}
respectively. The momentum equation (\ref{ns}) gives the rate of the energy dissipation
\begin{equation}\label{endiss}
	\frac{\rd E}{\rd t}=- \nu \:\Omega.
\end{equation}
Thus the principle of energy conservation is expressed in 
\begin{equation} \label{engy}
	E(t) + \nu \int_0^t \! \Omega(\tau) \:\rd \tau = E(0).
\end{equation}
Because of the term $\nabla{\times}\zeta$ in (\ref{ns}), we must examine the function palinstropgy
\begin{equation} \label{palin}
	Z(t)=\int_{\Upo} \Big( \: \big( \partial_x \zeta\big)^2 + \big(\partial_y \zeta \big)^2 \: \Big)(\bfx,t) \; \rd \bfx,
\end{equation}
as it is related to the rate of change in the enstrophy
\begin{equation}\label{ents}
	\frac{\rd \Omega}{\rd t}= - 2 \nu \: Z + 2 \nu\: \Big( \big[\zeta \: \zeta_y\big]_{x=0,1} + \big[\zeta \: \zeta_x \big]_{y=0,1} \Big),
\end{equation}
where the square brackets in the last term refer to the wall values. In general, the size of $Z$ is much larger than the last term so that the enstrophy is being consumed by viscous effects over flow evolution. 

An anomalous energy dissipation is a mathematical argument which conjectures the rate of energy dissipation $\rd E/\rd t$ would be independent of viscosity $\nu$ when viscosity becomes vanishingly small, see Kato (1984). In particular, the anomaly is assumed to occur in thin boundary layers in the vicinity of solid surfaces. In other words, should there exist a flow in which $\Omega \propto \nu^{-1}$, the energy of the flow remains to be dissipated and thus is an inviscid process. There has been an expectation, largely unjustified, that the Euler equations ($\nu=0$ in (\ref{ns})) are capable of describing fluid motions, even for turbulence. 

The anomaly hypothesis is established on the agreement of the energy norm between viscous and inviscid equations in the limit $\nu \rightarrow 0$. The mathematical analyses of a possible anomalous dissipation might be valid over short time intervals. There are no particular physical reasons which categorically exclude such momentary instances; it is a matter of the detailed local vorticity dynamics. In a flow having an arbitrary initial condition, the conservation law (\ref{engy}) does not say that the energy inside the square must continuously decrease from the start $t=0$. In fact, vorticity theory and numerical experiment suggest that the critical quantity is not the energy but the di-vorticity. Whether the conjecture is genuine or not depends on demonstrating, at least, {\it both} $\Omega(t) \;and \; Z(t) \propto \nu^{-1}$ over a sufficiently long period of time. 
\section{Dipole topology and dynamics}
For the present demanding problem of dipole-wall interaction, the vorticity field will undergo drastic changes over $\Delta t$. The errors of the incorrectly-imposed wall vorticity will contaminate the numerical solutions. For the incompressible flows, many previous studies show that numerical meshes $\sim O(1/\nu)$ have to be used in order to properly resolve various fine-scale motions. Such stringent requirements seem method-independent. For instance, Clerxc \& Bruneau (2006) made use of a pseudo-spectral scheme and a finite difference approximation to model dipole collisions. They have found that mode-convergence could hardly be achieved by coarse grids even at low Reynolds numbers of few thousands. Since the solution of the vorticity equation is unique and regular, we have to accept the reality that fairly dense meshes are a must for satisfactory numerical simulations of fluids.
\subsection*{Twin-core dipoles}
\begin{figure}[ht] \centering
  {\includegraphics[keepaspectratio,height=9.25cm,width=9.25cm]{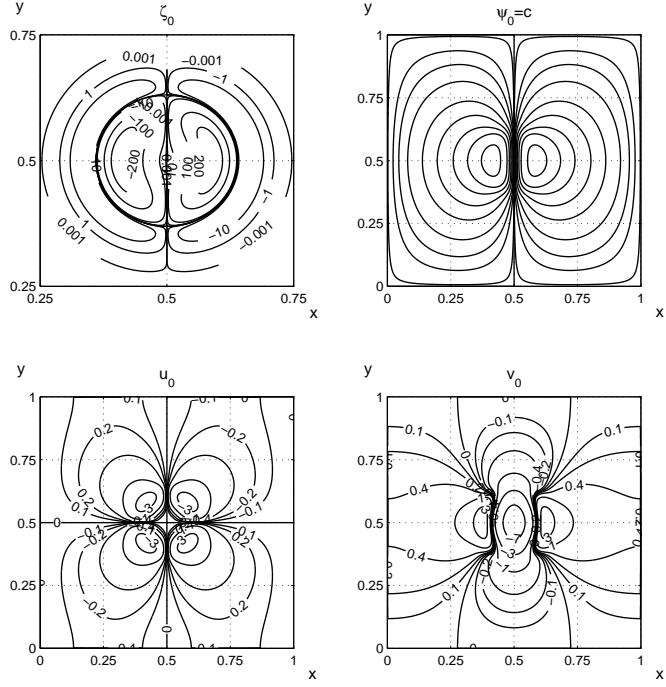}} 
 \caption{Dipole vortex (\ref{dipole}). The solenoidal velocity extends to the no-slip boundary walls. As the Poisson equation for the stream function is solved subject to the homogeneous boundary condition, the calculated velocities form closed loops next to the solid surfaces where viscous layers are formed according to the kinematics. The peak vorticity magnitude is $269.5$ at the centres of the vortices. Similarly, the maximum initial velocities are $7.96$ and $4.31$ for $u_0$ and $v_0$ respectively. The flow is indeed incompressible. } \label{dpole} 
\end{figure}

\begin{figure}[ht] \centering
  {\includegraphics[keepaspectratio,height=9.25cm,width=9.25cm]{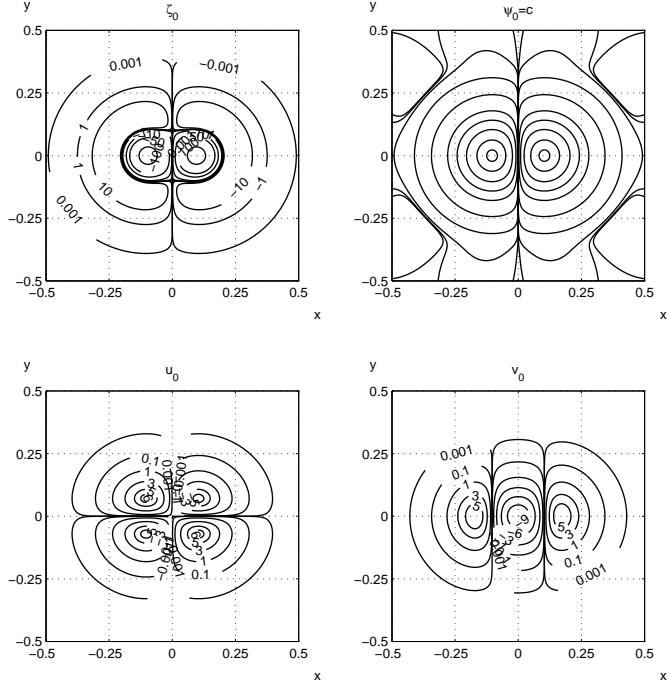}} 
 \caption{The core region of initial dipole vortex (\ref{chpole}) (Clercx \& Bruneau 2006). As the high shears are located in the centre, the initial boundary layers are negligible. This property is not important because, as soon as the vortices start to evolve, the no-slip condition will enforce the formation of wall viscous layers.} \label{chdipole} 
\end{figure}

A localised shear concentration at the centre of the square is given by
\begin{equation} \label{dipole}
	\zeta_0(\bfx) = 325\:\Big( \:\re^{-r_1^2} \; \big(1-r_1^2/b^2 \big) - \;\re^{-r_2^2} \big(1-r_2^2/b^2 \big) \:\Big),
\end{equation}
where $r_1=(2x{-}1{-}a)^2 + (2y{-}1)^2$, $r_2=(2x{-}1{+}a)^2 + (2y{-}1)^2$, $a=b=0.05$, and $0 \leq x,y\leq 1$ (see figure~\ref{dpole}). The constant $325$ is chosen so that the initial energy $E_0 \approx 1$ though this choice is not essential. 
Our initial vorticity dipole is derived from that of Clercx \& Bruneau (2006) or Kramer {\it et al}. (2007)
\begin{equation} \label{chpole}
	\zeta_0(\bfx) = 300\:\Big( \:\re^{-s_1^2} \; \big(1-s_1^2/0.01 \big) - \;\re^{-s_2^2} \big(1-s_2^2/0.01 \big) \:\Big),
\end{equation}
where $s_1^2=(x-0.1)^2+y^2$, $s_2^2=(x+0.1)^2+y^2$, for $-1 \leq x,y \leq 1$ (figure~\ref{chdipole}). 
The initial downward velocity at the core is greater than that of the preceding case.
\subsection*{Kirchhoff pair}
\begin{figure}[ht] \centering
  {\includegraphics[keepaspectratio,height=9.25cm,width=9.25cm]{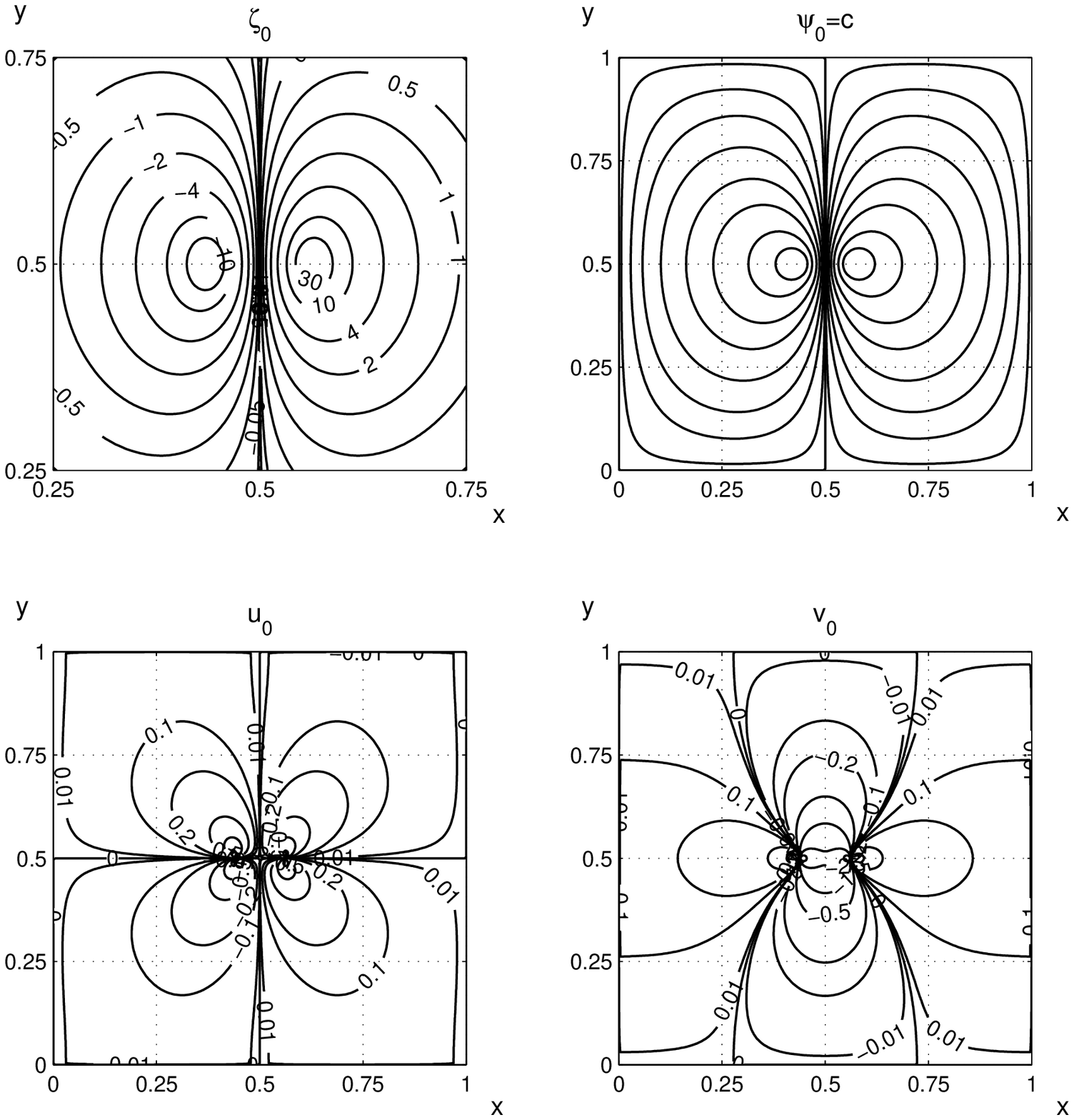}} 
 \caption{Initial data of Kirchhoff vortices (\ref{khoff}). Compared to (\ref{dipole}) in figure~\ref{dpole}, the present vorticity is much weaker with essential differences in the core structure.} \label{khoff0} 
\end{figure}

There are other simple functions that define vortex pairs. For example, the following algebraic expression resembles a couple of Kirchhoff vortices:
\begin{equation} \label{khoff}
\zeta_0(\bfx) = \exp(-r) \: \Big(\:\frac{x_f}{\kappa{+}(x_f{-}1/4)^2{+}y_f^2/2} + \frac{x_f}{\kappa{+}(x_f{+}1/4)^2{+}y_f^2/2}\:\Big),
\end{equation}
where $\kappa=10^{-4}$, $x_f=4x{-}2$, $y_f=4y{-}2$, and $r=x_f^2+y_f^2$, see figure~\ref{khoff0}.
\subsection*{Lamb dipole}
\begin{figure}[ht] \centering
  {\includegraphics[keepaspectratio,height=9.25cm,width=9.25cm]{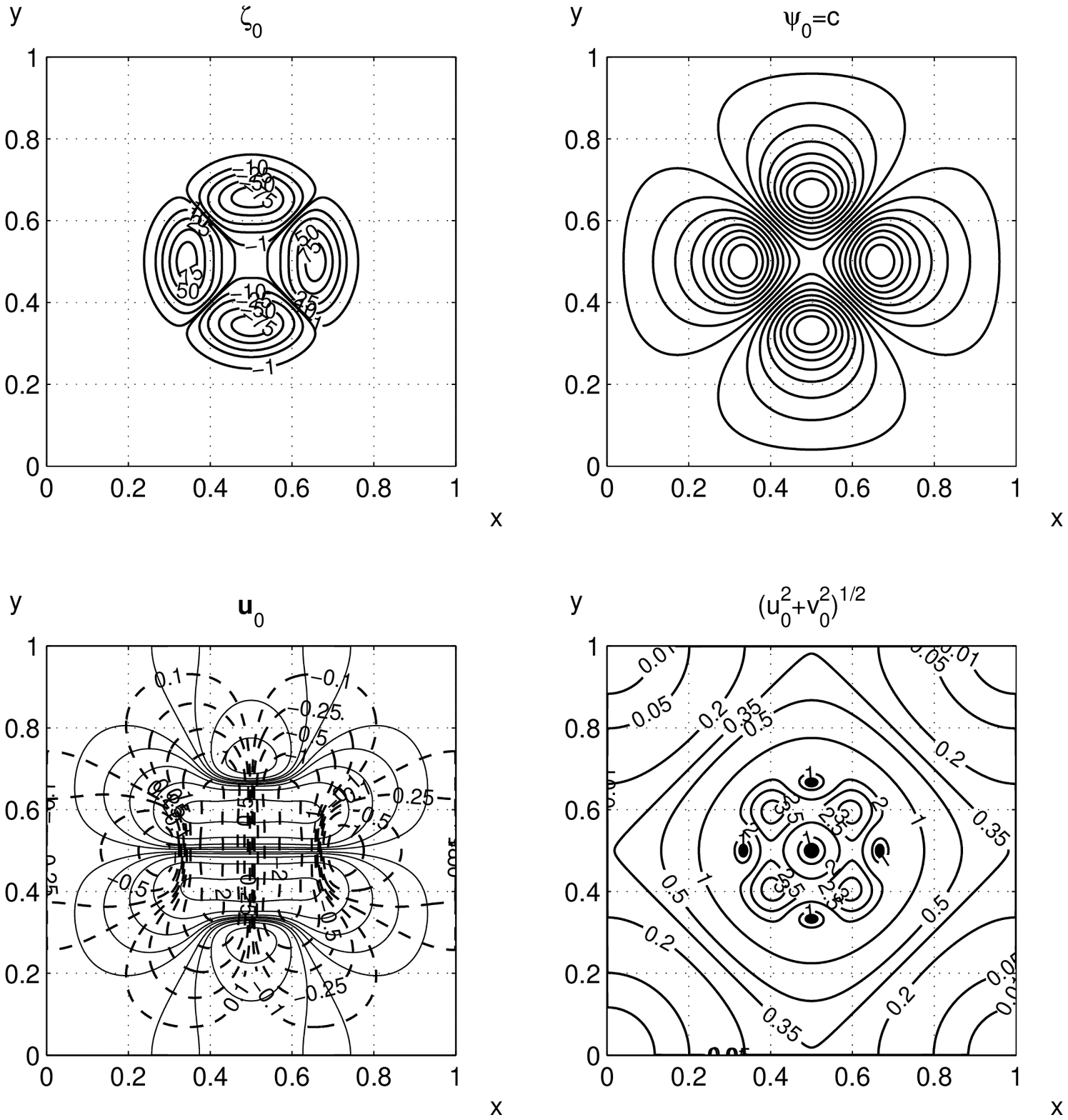}} 
 \caption{Symmetric dipole pair (\ref{lamb}) look simple and innocuous but their subsequent dynamics is a total surprise (see figure~\ref{lamb10k} for an example run at $\nu=10^{-4}$). In the velocity plot, solid lines are for $u_0$; dashed lines for $v_0$. } \label{lamb0} 
\end{figure}

The following exponential function gives a Lamb-dipole
\begin{equation} \label{lamb}
\zeta_0(\bfx) = (2\pi)^4\:r^2\:\exp\big( -2 \pi \: r^2 \big) \: \cos(2\theta)
\end{equation}
where $\theta=\tan^{-1}(y_f/x_f)$, see figure~\ref{lamb0}. The rotational symmetry parameter in the cosine function determines the number of vortex pairs.

\section{Discussion and outlook}
\begin{figure}[ht] \centering
  {\includegraphics[keepaspectratio,height=14cm,width=14cm]{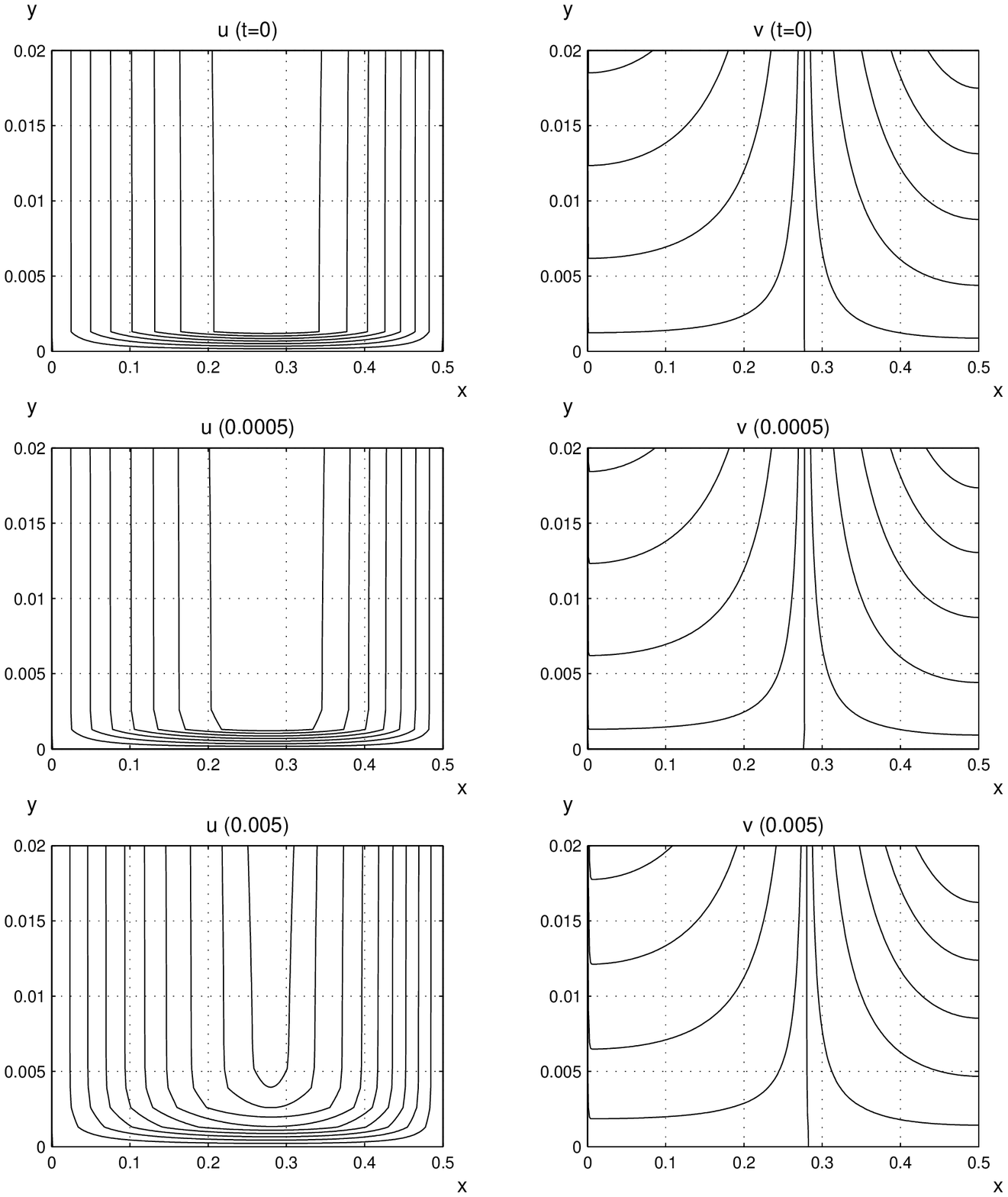}} 
 \caption{Immediate formation of viscous wall layers (\ref{dipole}), along lower wall $y=0$ and on side wall $x=0$. Computation at $\nu^{-1}=5000$, grid $768{\times}768$ and $\Delta t = 10^{-4}$. Plotted contours are $-0.17,-0.16,-0.14$,$-0.12,-0.1,-0.08$,$-0.06,-0.04$,$-0.02,0$ for $u$; $\pm0.02,\pm0.015,\pm0.01,\pm0.005,\pm0.001,0$ for $v$. Note that the wall viscous layers are determined by the vorticity dynamics and the no-slip boundary condition.} \label{uwall} 
\end{figure}

For the dipoles having certain symmetry, it is useful to examine the circulation inside the square 
\begin{equation*} 
	\Gamma(t)=\int_{\Upo} \zeta(\bfx,t) \;\rd \bfx = \int_{\bdy} \big( \; u \:\rd x + v \: \rd y \; \big) = 0.
\end{equation*}
There are no boundary conditions for the vorticity. As a definite routine, we choose to specify an arbitrary initial vorticity which may not produce a compatible velocity field. Thus the iterative method is first to solve the Poisson equation $\Delta \psi_0= -\zeta_0$ subject to the no-slip condition. Figure~\ref{uwall} illustrates how quickly the compatible solutions can be found. Within the viscosity range of the present note, the time increment $\Delta t$ is usually chosen as $O(10^{-4} \sim 10^{-5})$, then the numerical method finds the solenoidal velocity within $5$ to $10$ iterations. 

\begin{figure}[ht] \centering
  {\includegraphics[keepaspectratio,height=6cm,width=13cm]{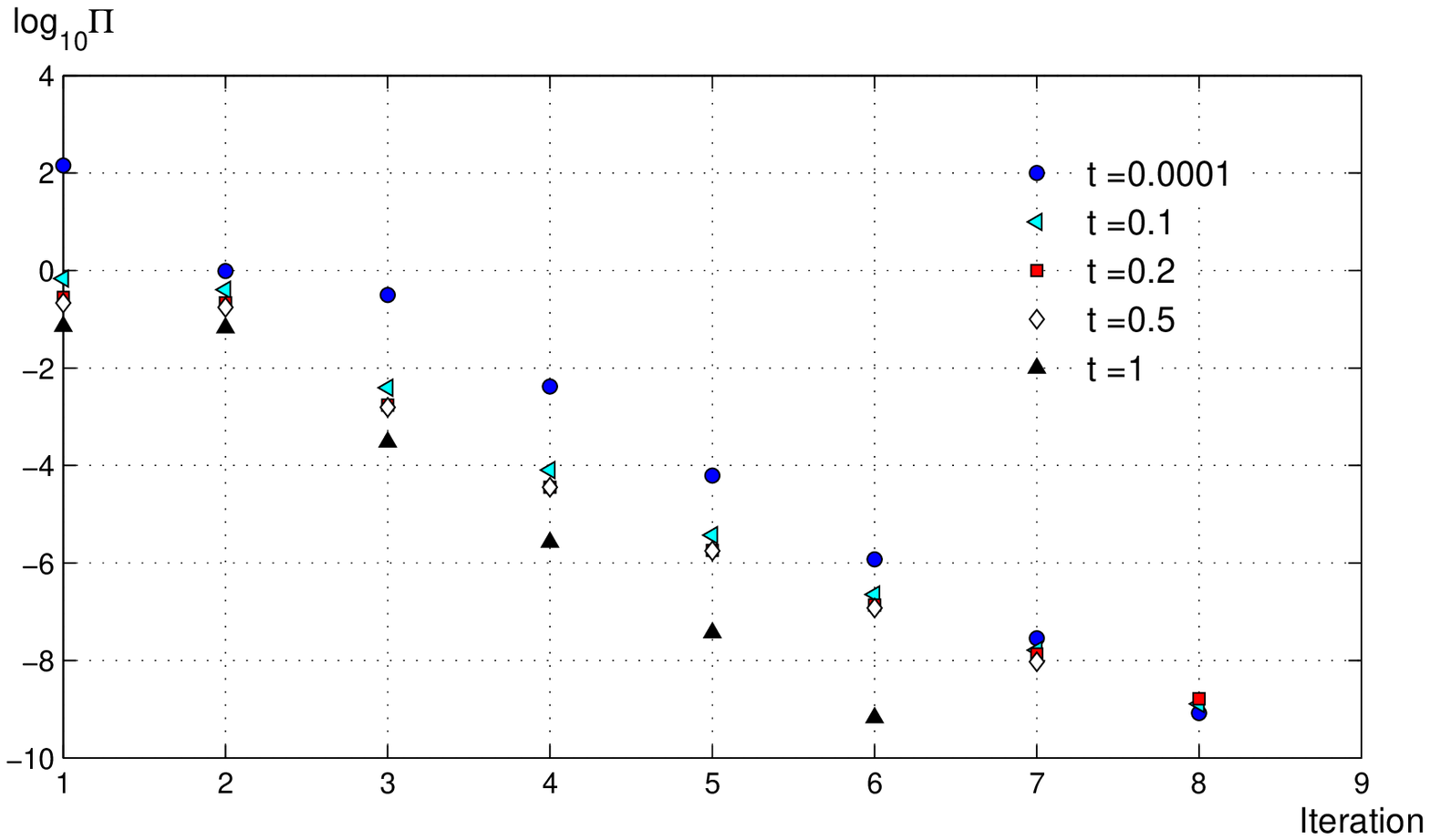}} 
 \caption{Examples of convergence rate in reducing the vorticity residue from data (\ref{chpole}) ($1/\nu=2500$, $\Delta t=10^{-4}$, grid $1536^2$). The first set of the data shows that the local converged solution and the initial data are ``far apart'' but the vorticity dynamics is satisfied by the formation of the wall layers. It is understandable that more iterations are required over the interval ($t<0.5$) where the interactions between the main vortices and the solid walls are substantial.} \label{cvgn2p5k} 
\end{figure}

\begin{figure}[ht] \centering
  {\includegraphics[keepaspectratio,height=11cm,width=11cm]{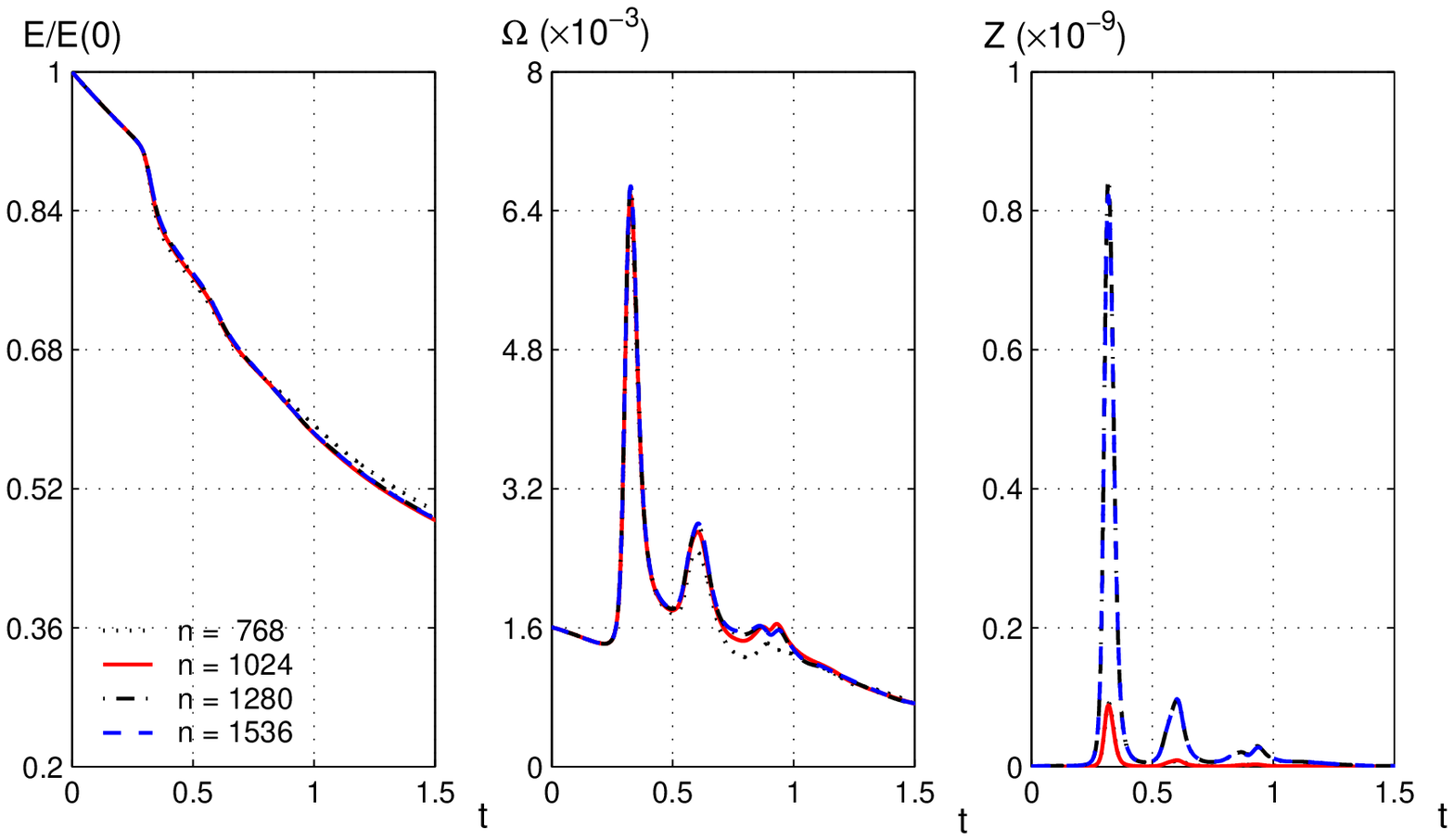}} 
 \caption{Good grid convergence for $n \geq 1280$ for test case (\ref{chpole}) of Clercx \& Bruneau (2006), $1/\nu=2500$ and marching time $\Delta t=10^{-4}$. The two lower mesh-grids ($768$ and $1024$) are inadequate in resolving the vorticity derivatives though the energy and enstrophy appear to have converged. As the equations of motion contain the Laplacian $\Delta \bfu= - \nabla{\times}\zeta$, it is not surprising that the local palinstrophy $Z(t)$ plays a key role in the determination of mesh resolution. At $t=1$, $|\zeta|_{\rm{max}} \approx 570$ which is much weaker than that at the second peak near $t \approx 0.606$ (cf. table~\ref{cmp}). However, there is an increase in $|u|$ from $5.2$ at $t=1$ to a peak $5.67$ at $t=1.018$, and the velocity attenuates thereafter. Our converged results should also be compared with Fig.4 of Kramer {\it et al}. (2007). } \label{meshcvgn} 
\end{figure}

\begin{figure}[t] \centering
  {\includegraphics[keepaspectratio,height=4.5cm,width=15cm]{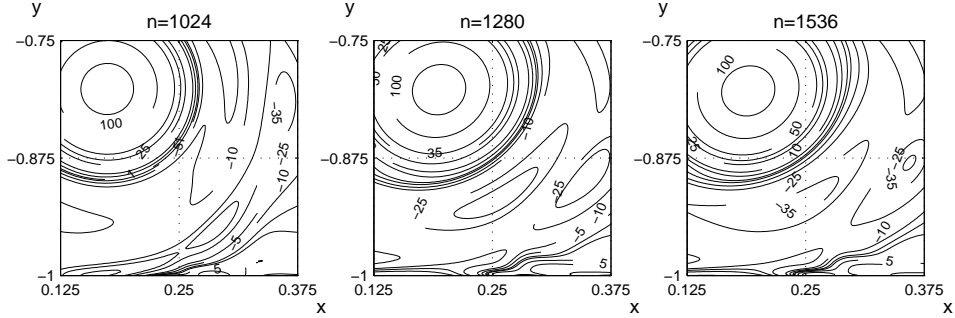}} 
 \caption{Vorticity solutions at $t=1.5$ show some minor differences among different meshes. Note that the truncation errors in our numerical scheme are $O(\Delta t)$ in time and $O(\Delta x^2)$ in space. Recall that all the derivatives are calculated by central differencing. Also there must be an accumulation of discretisation errors over time. The last two fine meshes are consistent within the numerical approximations.} \label{dztt1p5} 
\end{figure}

\begin{figure}[ht] \centering
  {\includegraphics[keepaspectratio,height=6cm,width=16cm]{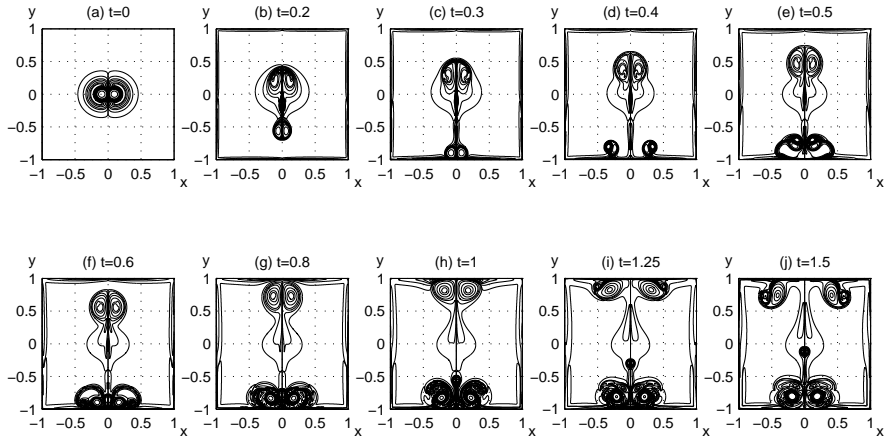}} 
 \caption{Vorticity development for test case (\ref{chpole}), $\nu^{-1}=2500$ and $1536$ grid points. Plotted contours are $\pm200, \pm100, \pm75, \pm50, \pm35, \pm30, \pm20, \pm10, \pm5, \pm1$ and $\pm0.01$. The secondary vortices are strong enough to influence the movement of the primary ones. } \label{chd2p5k} 
\end{figure}

Our numerical procedures are first validated against the results of Clercx \& Bruneau (2006) (see figure~\ref{chdipole}). In figure~\ref{cvgn2p5k}, we show one example that demonstrates the convergence of our iterative procedures. The results of mesh convergence and small-scale flow fields are given in figure~\ref{meshcvgn} to figure~\ref{dztt1p5}. The flow developments of dipole data (\ref{chpole}) are shown in figure~\ref{chd2p5k} for $Re=2500$. Tables~\ref{cmp} and \ref{cmpt5} list the numerical values of the comparison.

\begin{table}
	\centering
\begin{tabular}{c|cc|cc||cc|cc} \hline \hline
 & \footnotesize{$t_1$}	& \footnotesize{$\Omega_1$} &	\footnotesize{$t_2$}	& \footnotesize{$\Omega_2$}		& \footnotesize{$s_1$}	& \footnotesize{$Z_1$}	& \footnotesize{$s_2$}	 & \footnotesize{$Z_2$} \\ \hline \hline
\footnotesize{$1/\nu=625$} & & & & & & & &  \\ 								
\footnotesize{Present}	& \footnotesize{0.3702}	&  \footnotesize{940.7}	&  \footnotesize{0.6450}	&  \footnotesize{307.0}		&  \footnotesize{0.3622}	&  \footnotesize{2.574(7)}	&  \footnotesize{0.6504}	&  \footnotesize{1.317(6)} \\ 
\footnotesize{$(n{=}1024)$} & & & & & & & &  \\ 
\footnotesize{C-B}	& \footnotesize{0.3711} &	\footnotesize{933.6} &	\footnotesize{0.6479} &	\footnotesize{305.2} &		\footnotesize{0.3624} &	\footnotesize{2.772(7)} &	\footnotesize{0.6521} &	\footnotesize{1.355(6)} \\ \hline \hline
\footnotesize{$1/\nu=1250$} & & & & & & & &  \\ 
\footnotesize{Present}	& \footnotesize{0.3404}	& \footnotesize{1919.7}	& \footnotesize{0.6140}	& \footnotesize{728.1}		& \footnotesize{0.3317}	& \footnotesize{1.770(8)}	& \footnotesize{0.6199}	& \footnotesize{1.426(7)} \\
\footnotesize{$(n{=}1280)$} & & & & & & & &  \\ 
\footnotesize{C-B}	& \footnotesize{0.3414}	& \footnotesize{1899}	& \footnotesize{0.6162}	& \footnotesize{725.3}		& \footnotesize{0.3326}	& \footnotesize{1.742(8)}	& \footnotesize{0.6234}	& \footnotesize{1.432(7)} \\ \hline \hline
\footnotesize{$1/\nu=2500$} & & & & & & & &  \\ 
\footnotesize{Present}	& \footnotesize{0.3266}	& \footnotesize{3342.0}	& \footnotesize{0.6059}	& \footnotesize{1400.2}	&	\footnotesize{0.3184}	& \footnotesize{8.220(8)}	& \footnotesize{0.6022}	& \footnotesize{9.731(7)} \\
\footnotesize{$(n{=}1536)$} & & & & & & & &  \\ 
\footnotesize{C-B}	& \footnotesize{0.3279}	& \footnotesize{3313}	& \footnotesize{0.6089}	& \footnotesize{1418}		& \footnotesize{0.3195}	& \footnotesize{7.936(8)}	& \footnotesize{0.6046}	& \footnotesize{1.004(8)} \\ \hline \hline
\footnotesize{$1/\nu=5000$} & & & & & & & &  \\ 
\footnotesize{Present}	& \footnotesize{0.3223}	& \footnotesize{5543.4}	& \footnotesize{0.6039}	& \footnotesize{3715.7}		& \footnotesize{0.3210}	& \footnotesize{3.442(9)}	& \footnotesize{0.5994}	& \footnotesize{1.990(9)} \\
\footnotesize{$(n{=}2560)$} & & & & & & & &  \\ 
\footnotesize{C-B}	& \footnotesize{0.3234}	& \footnotesize{5536}	& \footnotesize{0.6035}	& \footnotesize{3733}		& \footnotesize{0.3219}	& \footnotesize{3.556(9)}	& \footnotesize{0.5992}	& \footnotesize{2.080(9)} \\ \hline \hline
	\end{tabular}
\caption{\rm{Comparison with the calculations of Clercx \& Bruneau (2006) (spectral method). The present time increment $\Delta t = 10^{-4}$ for all cases. Despite of the different methods of solution and their numerical implementations, the agreement is entirely satisfactory for both the impact time and enstrophy. Note that non-linear quantity $\nabla \zeta$ forms part of computations in the present iterative procedures. Nevertheless, the maximum relative errors in the palinstrophy are less than a few percent.}} \label{cmp}
\end{table}								

\begin{figure}[ht] \centering
  {\includegraphics[keepaspectratio,height=11cm,width=12.5cm]{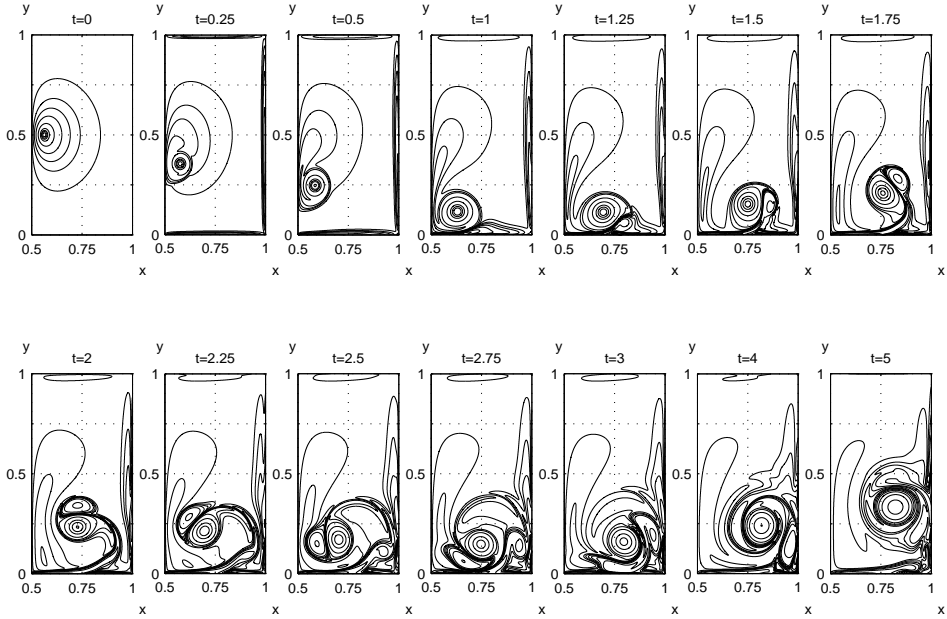}} 
 \caption{Wall impact of Kirchhoff vortices (\ref{khoff}). Computations at $\nu=10^{-4}$, grid $1024^2$, and $\Delta t = 10^{-4}$. Plotted contours are $\pm250, \pm100, \pm60, \pm30, \pm10, \pm4, \pm2, \pm1$ and $\pm0.25$. The initial field does not produce a pair of secondary eddies. The early development of the vortex structures (up to $t=2$) agrees with the anatomy of vortex-wall impingement as well-described by Orlandi (1990). Soon afterwards, the side no-slip wall forms a base for the main eddy to roll upward. There are no trailing jet pair. Naturally, this roll-up process cannot be sustained on periodic domains. } \label{khoff10k} 
\end{figure}

\begin{figure}[ht] \centering
  {\includegraphics[keepaspectratio,height=5cm,width=17cm]{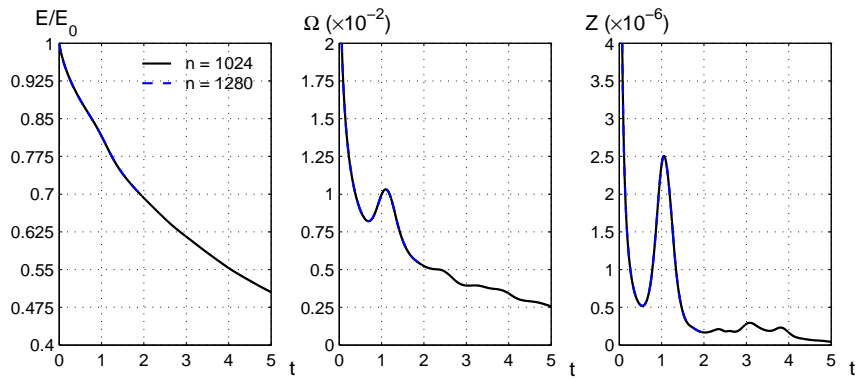}} 
 \caption{Computed integral properties of Kirchhoff dipole (\ref{khoff}) show that the wall impact is rather mild. A re-run with mesh $n=1280$ up to $t=2$ confirms the expectation that meshes in $O(10^3)$ are adequate to resolve the fine-scale motion. } \label{khoffcv} 
\end{figure}

\begin{figure}[ht] \centering
  {\includegraphics[keepaspectratio,height=12.5cm,width=12.5cm]{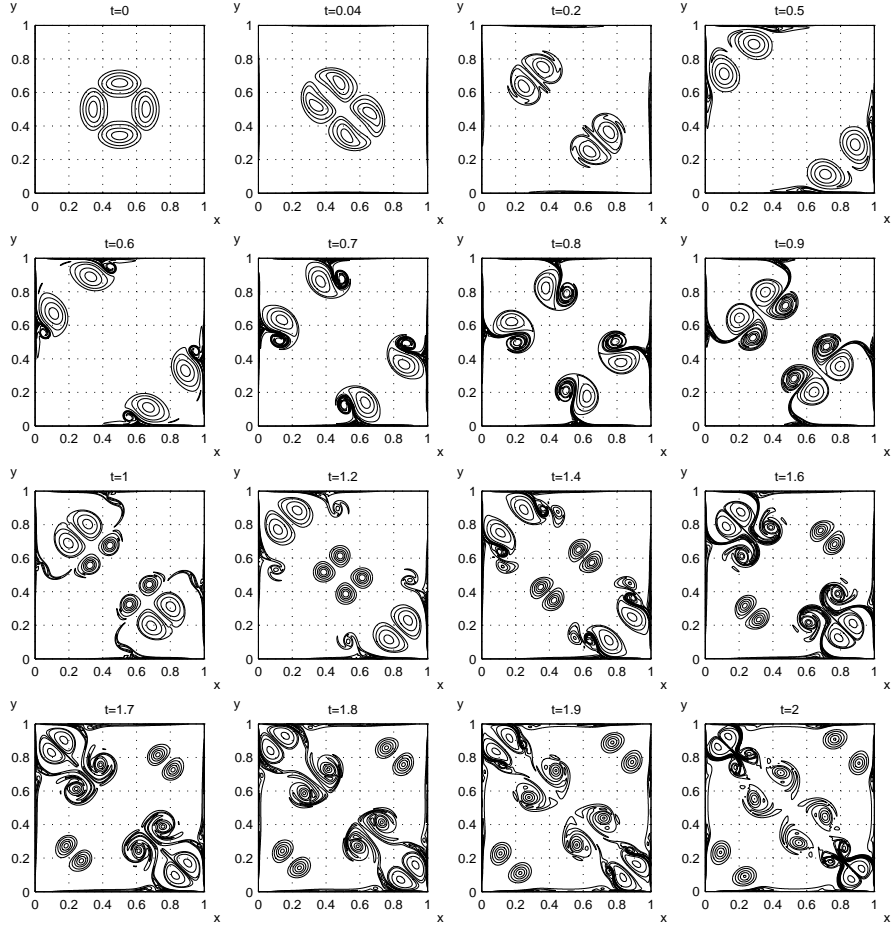}} 
 \caption{Lamb vortices (\ref{lamb}) simulated at grid $1024^2$ and $\Delta t = 10^{-4}$. Iso-contours are shown at $\pm100, \pm75, \pm50, \pm25$ and $\pm10$. Two eddy pairs of weaker shears are detached from the primary vortices near $t=1$. They are pushed and distorted in the main vorticity field before shooting off along the opposite diagonal. A series of highly deformed dipoles are borne of the bouncing primary pair. } \label{lamb10k} 
\end{figure}

\begin{figure}[ht] \centering
  {\includegraphics[keepaspectratio,height=4.5cm,width=16cm]{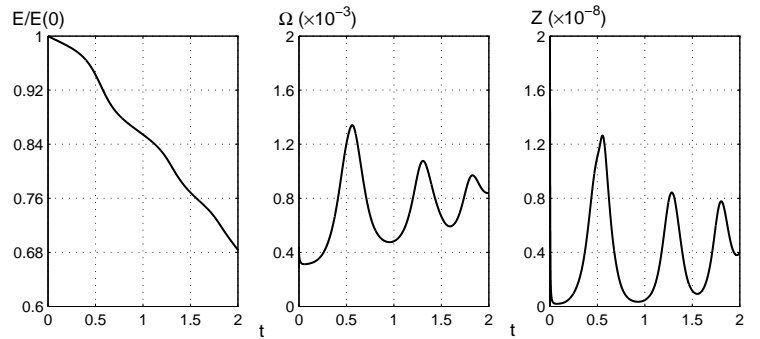}} 
 \caption{The wall impacts are relatively gentle with no ``sharp and narrow'' upsurges. } \label{lambhist} 
\end{figure}

The Kirchhoff dipole of simple algebraic description (\ref{khoff}) undergoes a much milder collision, see figure~\ref{khoff10k} and figure~\ref{khoffcv}. The evolution of the Lamb dipole (\ref{lamb}) is summarised in figure~\ref{lamb10k} and figure~\ref{lambhist}.

\begin{figure}[ht] \centering
  {\includegraphics[keepaspectratio,height=11cm,width=12.5cm]{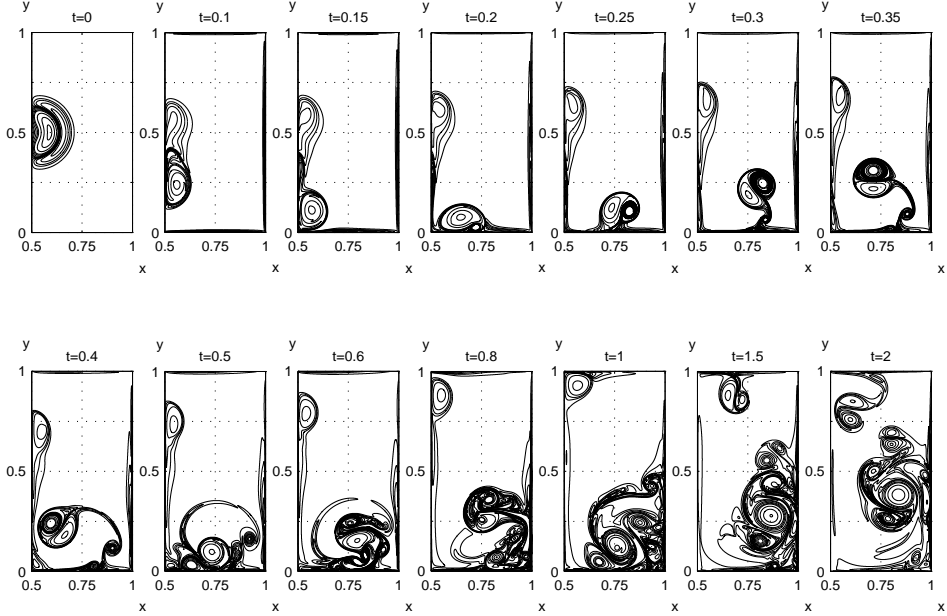}} 
 \caption{The flow evolution is marked by repeated production of mushroom vortices as a result of the interaction between the eddies of asymmetric shears and the no-slip walls (data (\ref{dipole}), $\nu=10^{-4}$, grid $2048{\times}2048$, and $\Delta t = 5{\times}10^{-5}$). Vorticity contours are $\pm250, \pm175$, $\pm100,\pm50,\pm35,\pm20,\pm10,\pm5$ and $\pm1$. The initial velocity indicates that there are $4$ separate cores at the vortex centre which produce a pair of secondary vortices that shot upward, rebound at the upper wall, and amalgamate into the primary vortical stream. The following plot shows the development history up to $t=2.5$. } \label{dp10k} 
\end{figure}

\begin{figure}[ht] \centering
  {\includegraphics[keepaspectratio,height=5cm,width=17cm]{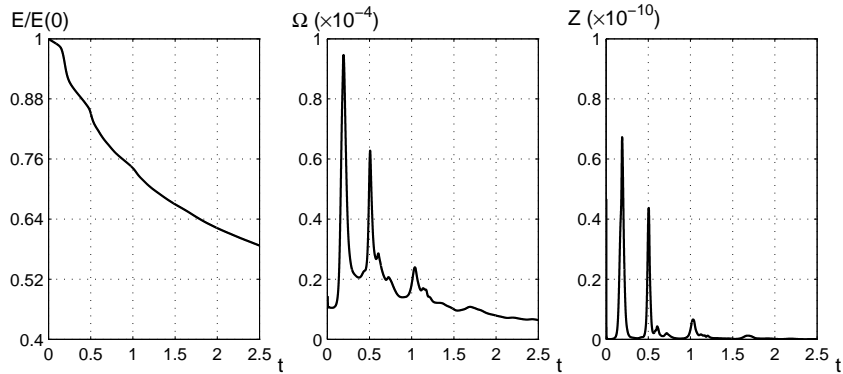}} 
 \caption{The surge in enstrophy at $t \approx 0.2$ records the main dipole-wall impact while the rebound near $t=0.5$ marks the re-impingement of the rolled vortices on the lower wall. The shell of the starting vortex travels upward and hits the upper wall at $t \approx 1$, and the impact is relatively weak. There are several mild shear-wall interactions over the interval $0.5 < t < 1.5$. } \label{dp10khist} 
\end{figure}

\begin{figure}[ht] \centering
  {\includegraphics[keepaspectratio,height=16cm,width=17cm]{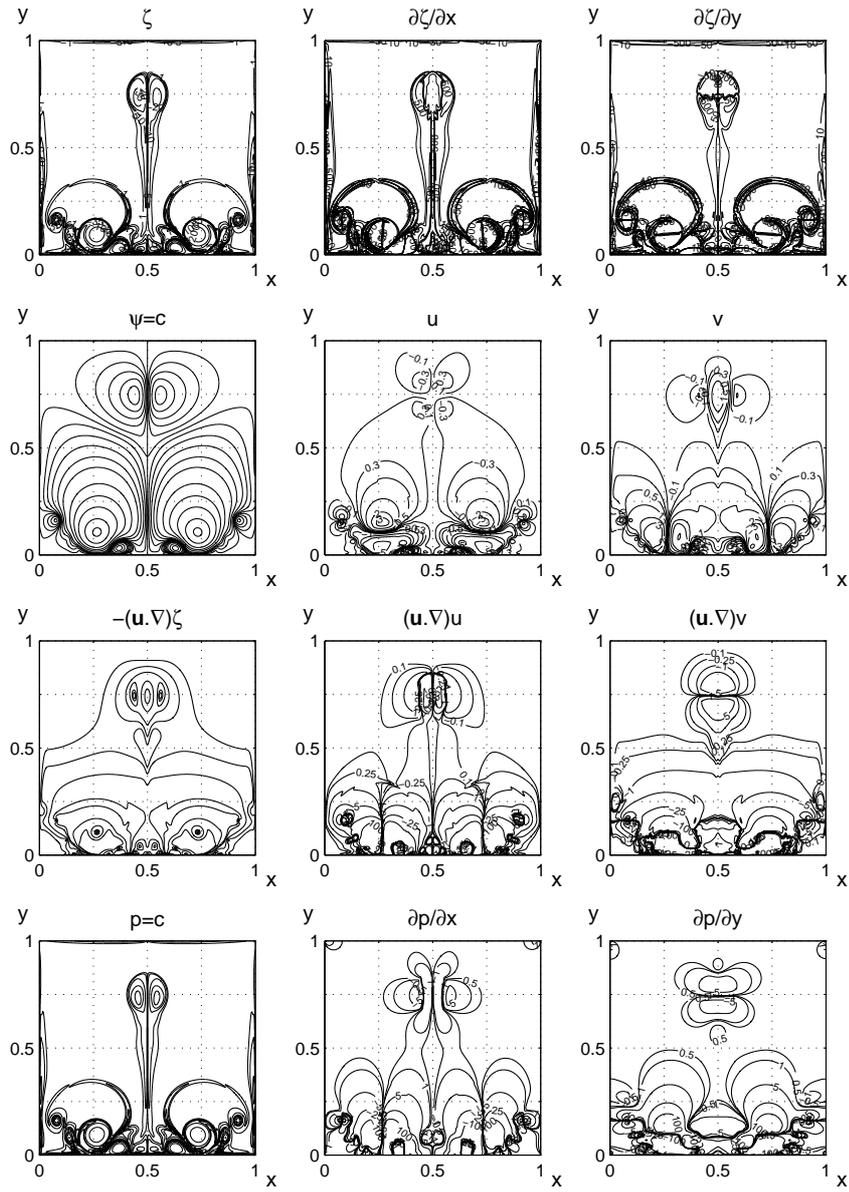}} 
 \caption{Detailed solutions are plotted at $t=0.5$ ($\nu=10^{-4}$). } \label{dp10k0p5} 
\end{figure}

\begin{figure}[ht] \centering
  {\includegraphics[keepaspectratio,height=12cm,width=15cm]{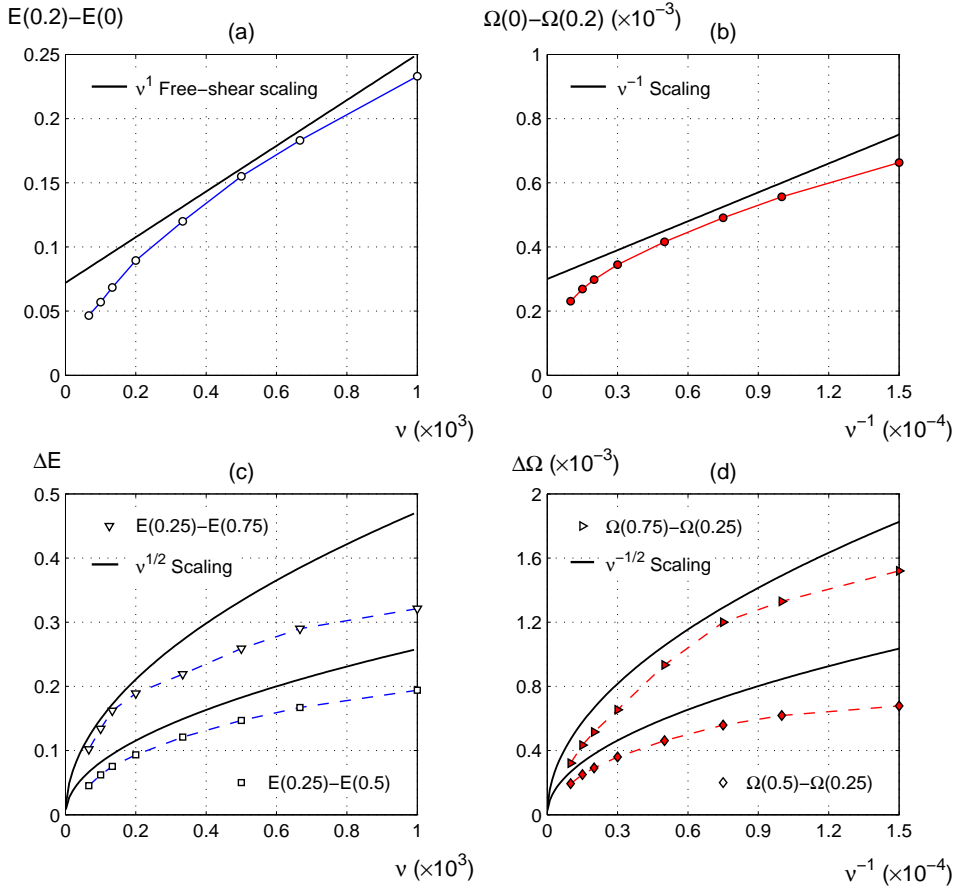}} 
 \caption{Viscosity scale of energy dissipation. The symbols are the present numerical computations and the solid lines refer to scaling laws. Plots (a) and (b) show the energy deficit and difference in the accumulative enstrophy before the dipole-wall impact. Our results recover the free-shear scaling, $E(t) \propto \nu$. In addition, the energy decay for small $\nu$ behaves like $\propto \nu^{\alpha}$, where $0<\alpha<1$, see (a). Thus the accumulative $\Omega$ or the second term on the left in (\ref{engy}) is strictly decreasing in the vanishing viscosity limit. Over the time intervals shown in (c) and (d), the flow undergoes significant non-linear evolution, giving rise to complicated structures in energy distribution (cf. figure~\ref{dp10khist}). Clearly, the Prandtl's boundary layer scaling ($\propto \sqrt{\nu}$) is fully applicable. Because of high quality of the numerical data, the $\nu$-scaling trends may be extrapolated to smaller values of viscosity. Throughout the present work, there has been no evidence that supports any persistence of the anomalous energy dissipation.} \label{dpeng} 
\end{figure}

As our revised initial dipole (\ref{dipole}) has a modest initial strength, its evolution is relatively easy to compute. We present a set of snap-shots in figure~\ref{dp10k} at $\nu=10^{-4}$. Figure~\ref{dp10khist} displays the variations of the integral quantities over time. It is hardly surprising to see the complexity of the detailed flow solution (figure~\ref{dp10k0p5}). Figure~\ref{dpeng} summarises the characters of the energy dissipation. Briefly, the conclusion of the present calculations is in line with the opinions of Sutherland {\it et al}. (2013).

The numerical formulation of the vorticity and stream-function has been further validated. We find that the iterative method is robust in the simulations of vortex-wall interaction. In practice, it is essential to ensure adequate temporal and spatial resolutions of the flow fields so as to avoid spurious solutions. 

We must admit that the notion of anomalous energy dissipation is obscured in physics. In brief, viscous effects, as dominated by fluids' microscopic structures, instigate vorticity gradients that are responsible for the energy degradation. There are other successful continuum systems, such as diffusion and heat transfer, that are formulated on the macroscopic scale on the premise of averaged microscopic contributions. 

The present solutions of vorticity dynamics clearly show that the unsteady separation of viscous-layer from the walls of the square is a consequence of the non-linearity in the full Navier-Stokes equation, characterised by $(\bfu.\nabla) \zeta$. Specifically, the pressure merely plays an auxiliary role in the dynamics. Large vorticity gradients can exist not only within the wall layers but also in the regions away square's boundaries. The integral (\ref{palin}) effects on di-vorticity $\nabla \zeta$ of the attached as well as the separated shears. With the advent of modern computational techniques, we should avoid the use of the boundary layer approximations. Versatile numerical solutions are bound to be instrumental to our understanding of complex flows at arbitrarily small viscosity. 

\vspace{1mm}
\begin{acknowledgements}
\noindent 
14 March 2019

\noindent 
\texttt{f.lam11@yahoo.com}
\end{acknowledgements}
%
%
\begin{figure}[t] \centering
  {\includegraphics[keepaspectratio,height=4.5cm,width=15cm]{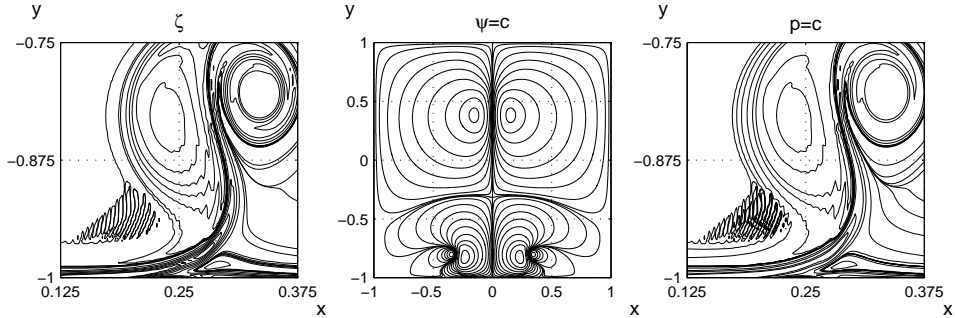}} 
 \caption{Flow solutions at $t=0.4$ and $1/\nu=2500$ with mesh of $768{\times}768$. Iso-vorticity values are $\pm285,\pm250,\pm200,\pm150,\pm100,\pm50,\pm10,\pm5$ and $\pm0.01$. The development of the contour ripples is due to insufficient spatial resolution though the current discretisation appears perfectly adequate for the stream function (the middle plot). Nevertheless, the velocities would appear to be well-resolved as $\psi$ has been smoothed in the Poisson regulator. On the other hand, the pressure field has been contaminated by the shear components of $\zeta$. The {\it oscillations} of small amplitude and high frequency are an artefact of inadequate numerics as the non-linear growth has not been properly accounted for. The poor approximations also produce {\it localised stitches} in iso-contours; these resulting shear patches must be understood as spurious flow scales.  } \label{dpripple} 
\end{figure}
\begin{figure}[t] \centering
  {\includegraphics[keepaspectratio,height=4.5cm,width=15cm]{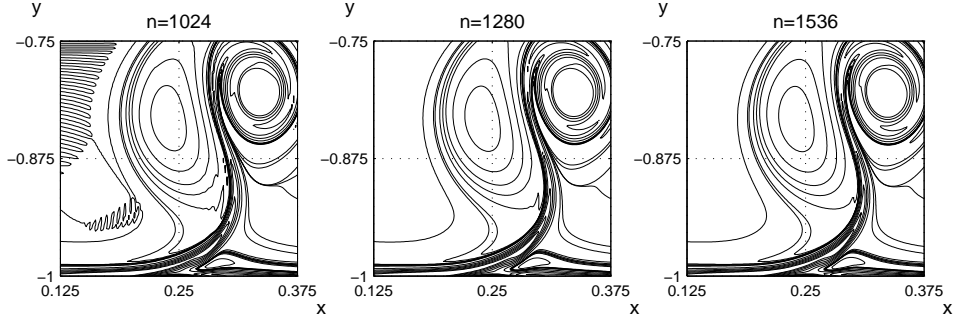}} 
 \caption{Flow solutions at $t{=}0.4$ and $1/\nu{=}2500$. Identical iso-vorticity contours as in figure~\ref{dpripple} with the addition of zero contour $\zeta{=}0$. As the flow-field resolution is progressively refined, the numerical snags are simply eliminated altogether. In practice, scale refinement and mesh-convergence check ought to be exercised in every simulation where intricate vorticity evolution is expected, because the non-linearity must be properly handled in numerical schemes. The above snapshots are selected from the solutions of the complete Navier-Stokes equations that offer the most comprehensive explanations of the fluid physics involved ($\nu > 0$), compared to the approximated theory by linearisation. In fact, the whole flow field is unsteady in nature and does not define any time-independent non-zero mean flow on which {\it ad hoc} perturbations may be superimposed. Consequently, the present example, like many other careful computations in the technical literature, substantiates the belief that the inviscid Rayleigh instability is irrelevant to the dynamics of the vortex-wall interactions. } \label{cvripple} 
\end{figure}
\begin{figure}[t] \centering
  {\includegraphics[keepaspectratio,height=4.5cm,width=15cm]{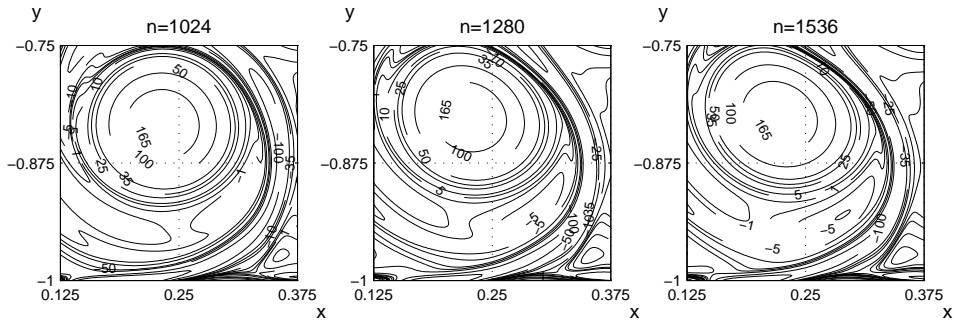}} 
 \caption{Flow solutions at $t=1$ and $1/\nu=2500$. The two finer meshes give almost identical results. (The tool for contour plotting appears to depend numerically on the fineness of the mesh grid.) Clearly, the wall derivatives of the coarsest mesh are poorly-defined. In part, the comparison explains the inconsistency of the enstrophy dissipation highlighted in figure~\ref{meshcvgn}. } \label{dztt1} 
\end{figure}
\newpage
\begin{table}
	\centering
\begin{tabular}{c|ccc||ccc} \hline \hline
 Time $t$	& $E$ &	$\Omega$	& $Z$		& $E$	& $\Omega$	& $Z$	 \\ \hline 			
 &  & (Present)	&  &  & (Clercx-Bruneau) &  \\ \hline
0 & 2.0062 &	1605.0	& 8.86(5) & 2	& 1600.0	& 8.84(5) \\
0.25 &	1.8568& 1464.0	& 8.65(6) & 1.8509 & 	1456.4	& 8.44(6) \\
0.5 &	1.5407 &	1826.7	& 7.76(6) & 1.5416	& 1841.0	& 7.66(6) \\
0.75	& 1.3268&	1575.8	& 7.15(6) & 1.3262	& 1616.2	& 7.58(6) \\ \hline \hline
	\end{tabular}
\caption{\rm{Comparison with Table 5 of Clercx \& Bruneau (2006) (spectral method). $Re=2500$. Satisfactory agreement is found in the different numerical schemes.}} \label{cmpt5}
\end{table}								
%

%
\newpage
%
%
\newpage
%
\newpage

\end{document}